\documentclass[10pt,twocolumn,superscriptaddress,nofootinbib,aps,floatfix,amsfonts,preprintnumbers,amsmath,amssymb,groupedaddress,prd]{revtex4-1}

\pdfoutput=1

\usepackage{graphicx}
\usepackage[colorlinks,pdfusetitle]{hyperref}

\usepackage{color}

\newcommand{\orcid}[1]{\href{https://orcid.org/#1}{#1}}
\newcommand{\wh}[1]{\widehat{#1}}
\newcommand{\h}{\mathcal H}

\newcommand{\Dmsqee}{\Delta m^2_{ee}}

\newcommand{\Szs}{\mathcal Z}
\newcommand{\Ssol}{\mathcal R_{\odot}}
\newcommand{\Satm}{\mathcal R_{\rm atm}}

\begin{document}


\title{Interplay between the factorization of the Jarlskog Invariant\\ and location of the Solar and Atmospheric Resonances \\  for Neutrino Oscillations in Matter
}

\author{Stephen J.~Parke}
\email{parke@fnal.gov}
\thanks{orcid \# \orcid{0000-0003-2028-6782}}
\affiliation{Theoretical Physics Dept., Fermi National Accelerator Laboratory, Batavia, IL 60510, USA}

\begin{abstract}

The Jarlskog invariant which controls the size of intrinsic CP violation in neutrino oscillation appearance experiments is modified by Wolfenstein matter effects for neutrinos propagating in matter.
 In this paper we give the exact factorization of Jarlskog invariant in matter into
the vacuum Jarlskog invariant times two, two-flavor matter resonance factors that control the matter effects for the solar and atmospheric resonances independently.   We compare the location of the minima of the factorizing resonance factors with the location of the solar and atmospheric resonances, precisely defined.  They are not identical but the fractional differences are both found to be less than 0.1\%.  In addition, we explain why symmetry polynomials of the square of the mass of the neutrino eigenvalues in matter,  such as inverse of the square of the Jarlskog invariant in matter, can be given as  polynomials in the matter potential.
\end{abstract}

\preprint{FERMILAB-PUB-20-609-T}

\date{December 14, 2020}

\maketitle

\section{Introduction}
\label{sec:introduction}

In the current and future long baseline experiments,  T2K \cite{Itow:2001ee}, NOvA \cite{Ayres:2004js}, DUNE \cite{Acciarri:2016crz} and T2HK(K) \cite{Abe:2014oxa,Abe:2016ero}, neutrinos propagate  between 300 to 1300 km through the earth's crust between source and detector.  Neutrino oscillations in matter are significantly different than in vacuum because of the coherent forward scattering of the neutrinos off the electrons in matter. This effect is known as the Wolfenstein matter effect  \cite{Wolfenstein:1977ue}.

The neutrino mass eigenvalues and the mixing angles of the neutrinos in matter are strongly impacted by the matter effect which depends  on density of the matter and the energy of the neutrino  through the Wolfenstein matter potential \cite{Wolfenstein:1977ue}, $a$, given by
\begin{equation}
a \equiv 2\sqrt{2} G_F N_e E_\nu\,,
\end{equation}
where  $G_F$ is the Fermi constant, $N_e$ is the number density of electrons, and $E_\nu$ is the neutrino energy in the matter rest frame.
These changes in the masses and the mixing angles have a significant effect on the oscillation probabilities as the neutrinos transverse through matter.

One of the primary goal of the above long baseline accelerator experiments is to measure intrinsic CP violation in the neutrinos sector. In vacuum, the CP-violating part of the neutrino oscillation probability in the appearance channels, e.g.~$\nu_\mu \rightarrow \nu_e$, is given by \cite{Bilenky:1987ty}
\begin{equation}
8 J\sin \Delta_{31} \sin \Delta_{32} \sin \Delta_{21}\,,
\label{eq:CPvac}
\end{equation}
where the kinematic phases are given by $\Delta_{jk} = \Delta m^2_{jk} L/4E_\nu$ with $\Delta m^2_{jk} = m^2_j -m^2_k$ for an experiment of baseline $L$ and neutrino energy $E_\nu$.
$J$ is the Jarlskog invariant \cite{Jarlskog:1985ht}, that controls the size of CP violation. Using the standard parameterization of the  Pontecorvo, Maki, Nakagawa and Sakata (PMNS) matrix \cite{Maki:1962mu,Pontecorvo:1967fh}, the Jarlskog invariant
 is given by
\begin{equation}
J\equiv s_{23}c_{23} s_{13}c^2_{13}s_{12}c_{12}  \sin\delta\,,
\label{eq:Jvac}
\end{equation}
where $s_{ij}=\sin\theta_{ij}$, $c_{ij}=\cos\theta_{ij}$.

For neutrinos propagating in matter,
the part of the appearance  oscillation probability that depends on the intrinsic CP violation is given by
\begin{equation}
8 \wh{J}\sin \wh{\Delta}_{31} \sin \wh{\Delta}_{32} \sin \wh{\Delta}_{21}\,,
\label{eq:CPmat}
\end{equation}
where $\wh{x}$ is the matter value for the vacuum variable $x$.
The Jarlskog invariant in matter, $\wh{J}$, is given by same expression as eq.~\ref{eq:Jvac}, but with the mixing angles and phase replaced by their matter values \cite{Zaglauer:1988gz,Krastev:1988yu,Parke:2000hu}.
The variables $\wh{\theta}_{12}$, $\wh{\theta}_{13}$ and $\Delta \wh{m^2}_{jk}$, that appear in eq.~\ref{eq:CPmat},
 have a strong dependence on  the Wolfenstein matter potential, $a$.

In \cite{Denton:2019yiw} by Denton and Parke (DP),  a simple and precise factorization of the Jarlskog invariant in matter was given as follows:
\begin{equation}
\wh{J} \approx \frac{J}{\Ssol^\text{appx}\, \Satm^\text{appx}}\,,
\label{eq:Jmat}
\end{equation}where
\begin{align}
\Ssol^\text{appx} &=\sqrt{1-2c_{13}^2\cos2\theta_{12}(a/\Delta m^2_{21})+c_{13}^4(a/\Delta m^2_{21})^2 }\,, \nonumber \\
\Satm^\text{appx} &=\sqrt{1-2 \cos2\theta_{13}(a/\Dmsqee)+(a/\Dmsqee)^2}\, .
\label{eq:S13,S12}
\end{align}
The fractional precision of this factorization is better than 0.07\%, for all $a$ and both mass orderings.  A factorization with a precision of a few per cent was given in \cite{Wang:2019yfp} with a follow up paper,  \cite{Wang:2019dal}, that reproduced the precise factorization of DP.
In this paper we give an extension of the analysis of DP to derive an exact factorization of this Jarlskog invariant in matter as well as a perturbative expansion in the two small quantities $s^2_{13}$ and $\Delta m^2_{21}/\Delta m^2_{ee}$.


Then, after defining precisely and calculating the location of the solar and atmospheric resonances to the relevant accuracy in the matter potential for the first time, we compare the location of the resonance with the location of the minima of the factorizing two flavor resonance factors. 
Although they are identical at the order given by DP,  what is show here, is they are not identical at higher orders but the fractional differences are found to be small, less than 0.1\%.  A qualitative understanding of why they differ is also given.

As part of our analysis, we show the simplicity of variables that can be expressed as a symmetric polynomial of the neutrino eigenvalues in matter, $\wh{m^2}_j$, such as the Jarlskog invariant in matter as well as other examples, Section II. We give the exact factorization of the Jarlskog invariant in matter as a product of two, two flavor resonance factors, Section III as well as a perturbative expansion. In Section IV, we compare the minima of these exact factorizing factors  to the actual location of the solar and atmospheric resonance for the matter potential, followed by a summary and conclusion section.  There are five Appendices that contain computational details.  Thus this paper provides additional understanding of the Wolfenstein matter effect for three neutrinos propagating through matter.

\section{Simplicity of Symmetric Polynomials of the Eigenvalues}
In the flavor basis, the neutrino propagation Hamiltonian in matter  is given by
\begin{multline}
(2E) H= 
 U
\begin{pmatrix}
0\\&\Delta m^2_{21}\\&&\Delta m^2_{31}
\end{pmatrix}
U^\dagger +
\begin{pmatrix}
a&&\\&0\\&&0
\end{pmatrix}
\, .\label{eq:ham}
\end{multline}
$U$ is the PMNS \cite{Pontecorvo:1967fh,Maki:1962mu}  lepton mixing matrix, parameterized by
\begin{multline}
 U= U_{23}(\theta_{23},\delta) U_{13}(\theta_{13}) U_{12}(\theta_{12})\equiv \\
\begin{pmatrix}
1\\&c_{23}&s_{23}e^{i\delta}\\&-s_{23}e^{-i\delta}&c_{23}
\end{pmatrix} 
\! \! \!
\begin{pmatrix}
c_{13}&&s_{13}\\&1\\-s_{13}&&c_{13}
\end{pmatrix}
\! \! \!
\begin{pmatrix}
c_{12}&s_{12}\\
-s_{12}&c_{12}\\
&&1
\end{pmatrix},
\end{multline}
where $s_{ij}=\sin\theta_{ij}$ and $c_{ij}=\cos\theta_{ij}$  are given by
\begin{align}
\sin^2 \theta_{13}& \equiv  |U_{e3}|^2 \approx 0.022 \, ,  \notag \\
\sin^2 \theta_{12} &  \equiv  |U_{e2}|^2/(1-|U_{e3}|^2) \approx 0.32 \, , \label{eq:ssqs} \\
\sin^2 \theta_{23} & \equiv   |U_{\mu3}|^2/(1-|U_{e3}|^2) \approx 0.55 \, . \notag
\end{align}
The sine and cosine of $\delta$ are given by
\begin{align}
e^{i\delta} & \equiv   \frac{(1-|U_{e3}|^2) U_{e2} U^*_{\mu 2} U^*_{e3} U_{\mu 3}  +|U_{e2}|^2|U_{e3}|^2 |U_{\mu 3} |^2}
{|U_{e1} U_{e2} U_{e3} U_{\mu 3} U_{\tau 3}|}  \, . \label{eq:delta} 
\end{align}
At this time, circa 2020, the numerical value of $\delta$ is still to be determined.  
Our definition of the CP-violating phase $\delta$ is  invariant under re-phasing\footnote{The commonly used definition, $\delta = -\rm{Arg}(U_{e3})$, is not re-phasing invariant and requires an explicit phase choice especially in the $U_{\alpha j}$ elements,  for $\alpha=\mu$ and $\tau$, j=1 and 2.} of the rows and columns of $U$ and thus
shifting it from it's PDG location, i.e. from next to $s_{13}$ to next to $s_{23}$, does not affect any observable\footnote{The sign of $\delta$ given by eq.~\ref{eq:delta} is consistent with the sign in the PDG expression for PMNS matrix.}. This choice is made because the $\theta_{23}, \delta$ sub-matrix of U, commutes with the matter potential and factors these variables from many expressions. For other choices for the ordering of the factorization of the PMNS matrix see \cite{Denton:2020igp}.

For the long baseline neutrino oscillation experiments,  T2K, NOvA, DUNE  and T2HK(K) it is sufficient to consider the matter density along the path of the neutrino to be a constant, $a$, as has been discussed in detail in \cite{Kelly:2018kmb,King:2020ydu}. Therefore, for the rest of this paper we will consider the matter potential to be a constant for a neutrino of a given energy.  This simplifies the solution to the evolution of the neutrino state in matter significantly, by allowing for exact analytical solution. However, the resulting solution is still  analytically impenetrable as we will see later in this section.

The characteristic equation for the matrix (2E)H gives the eigenvalues 
of the square of the neutrino masses in matter, $\wh{m^2}_j$,  all satisfying
\begin{align}
\left(\wh{m^2}_j \right)^3 -A\,\left(\wh{m^2}_j\right)^2+B\, \wh{m^2}_j  -C=0\,,
\label{eq:char}
\end{align}
where $A$, $B$, and $C$ are the sum of the eigenvalues, sum of the products of the eigenvalues, and the triple product of the eigenvalues:
\begin{align}
A&\equiv\sum_j \wh{m^2}_j 
=\Delta m^2_{31}+\Delta m^2_{21}+a\,,\nonumber\\
B&\equiv\sum_{j>k}\wh{m^2}_j\wh{m^2}_k 
= \label{eq:ABC}\\
&\Delta m^2_{31}\Delta m^2_{21}+a(\Delta m^2_{31}c^2_{13}+\Delta m^2_{21}(c^2_{12}+s^2_{13}s^2_{12}))\,,\nonumber\\[2mm]
C&\equiv\prod_j \wh{m^2}_j 
=a \Delta m^2_{31} \Delta m^2_{21} c^2_{13} c^2_{12}\, . \nonumber
\end{align}
We use the convention that  in vacuum $(\wh{m^2}_1, \wh{m^2}_2, \wh{m^2}_3 
)=(0,\Delta m^2_{21}, \Delta m^2_{31})$.

From refs.~\cite{cardano}, \cite{Barger:1980tf}, \cite{Zaglauer:1988gz}, the exact eigenvalues in matter are,
\begin{align}
\wh{m^2}_1&=\frac13A-\frac13\sqrt{A^2-3B}\left(\Szs+\sqrt3\sqrt{1-\Szs^2}\right)\, , \notag \\
\wh{m^2}_2&=\frac13A-\frac13\sqrt{A^2-3B}\left(\Szs-\sqrt3\sqrt{1-\Szs^2}\right)\,, \label{eq:whmsq}\\
\wh{m^2}_3&=\frac13A+\frac23 \sqrt{A^2-3B}\,\Szs\,. \notag
\end{align}
Where $\Szs$ contains the $\cos\{\frac13\cos^{-1}[\cdots]\}$ terms, given by
\begin{align}
\Szs &=\cos\left\{\frac13\cos^{-1}\left[\frac{2A^3-9AB+27C}{2(A^2-3B)^{3/2}}\right]  +\zeta \right\}\, ,\label{eq:S}
\end{align}
with $\zeta=0$ for normal ordering (NO) that gives $\wh{m^2}_1<  \wh{m^2}_2  < \wh{m^2}_3$  and  $\zeta=2\pi/3$ for inverted ordering (IO) that gives $\wh{m^2}_3<  \wh{m^2}_1  < \wh{m^2}_2$, the usual conventions.
 This convoluted term,  $\cos\{\frac13\cos^{-1}[\cdots]\}$, which is a generic feature of the analytic solution to cubic equations, does not lend itself to a useful perturbative expansion for arbitrary values of the matter potential, even though, for neutrino oscillations there are two small parameters $\sin^2 \theta_{13} \sim 0.02$ and $\Delta m^2_{21}/\Delta m^2_{31} \sim 0.03$. This unfortunate fact\footnote{As an example of the analytic impenetrability of $\Szs$, setting $a=0$ and recovering the vacuum values for the eigenvalues, $(0,\Delta m^2_{21}, \Delta m^2_{31})$, is a highly non-trivial exercise.
In vacuum $\Szs =\frac12 (\Delta m^2_{31}+\Delta m^2_{32})/ \sqrt{(\Delta m^2_{32})^2 + \Delta m^2_{21}\Delta m^2_{31}}$ for NO.}
  makes the exact analytical solution, eq.~\ref{eq:whmsq}, only useful, in general, for numerical studies, not analytic understanding.

However, while the exact eigenvalues have a very complicated analytic form,
any symmetric polynomial in the eigenvalues can be uniquely expressed in terms of  A, B \& C. This follows from the Fundamental Theorem of Symmetric Polynomials (FTSP), \cite{wiki:poly2}, and allows any symmetric polynomial of the $\wh{m^2}_i$ to be written as a polynomial in the matter potential, substantially simplifying such expressions.

An example of this is given in \cite{Harrison:1999df}, \cite{Yokomakura:2000sv}  \& \cite{Denton:2019yiw},
\begin{multline}
\left(\prod_{j>k}\Delta\wh{m^2}_{jk}\right)^2=\\(A^2-4B)(B^2-4AC)+(2AB-27C)C\, ,
\label{eq:Dprod}
\end{multline}
which given the expressions for A, B \& C in eq.~\ref{eq:ABC} is a fourth order polynomial in the matter potential,  $a$. No  impenetrable 
$\cos\{\frac13\cos^{-1}[\cdots]\}$ terms appear.

The Eigenvector-Eigenvalue Identity,  \cite{Denton:2019ovn} \& \cite{Denton:2019pka},  gives us the elements of the PMNS matrix in matter as follows
\begin{equation}
|\wh U_{\alpha i}|^2=\frac{(\wh{m^2}_i)^2 -(\xi+\chi)_\alpha \wh{m^2}_i   +  (\xi \chi)_\alpha}{\Delta \wh{m^2}_{ij}\Delta \wh{m^2}_{ik}}\,,\label{eq:Uaisq}
\end{equation}
where $i$, $j$  and $k$ are all different. The variables $(\xi+\chi)_\alpha $ and $(\xi \chi)_\alpha$,  which are the trace and determinant, respectively, of (2E)H with the $\alpha$-row and $\alpha$-column removed.  All $(\xi+\chi)_\alpha $ and $(\xi \chi)_\alpha$  are linear polynomials in the matter potential, $a$, and are given in Appendix \ref{sec:xichi}.

Combining eq.~\ref{eq:Dprod} \& \ref{eq:Uaisq},  we have that
\begin{multline}
 \left(\prod_{j>k}\Delta\wh{m^2}_{jk} \right)^2 \   \left( \prod_i |\wh U_{\alpha i}|^2 \right) =\\ 
 \prod_i \left((\wh{m^2}_i)^2 -(\xi+\chi)_\alpha \wh{m^2}_i-(\xi \chi)_\alpha \right)
 \label{eq:master}
\end{multline}
is a symmetric polynomial in the eigenvalues, $ \wh{m^2}_i $, and thus can be written as a polynomial in A, B, C, $(\xi+\chi)_\alpha$ and  $(\xi \chi)_\alpha$, as follows
\begin{align}
& \prod_i \left((\wh{m^2}_i)^2 -(\xi+\chi)_\alpha \wh{m^2}_i+(\xi \chi)_\alpha \right) \notag \\
=  & ~~C[C-B(\xi+\chi)_\alpha+A(\xi+\chi)_\alpha^2-(\xi+\chi)_\alpha^3]  \\
  & -(\xi+\chi)_\alpha(\xi\chi)_\alpha [(AB-3C)-B(\xi+\chi)_\alpha+A(\xi\chi)_\alpha ]  \notag  \\
  &+(\xi\chi)_\alpha [(B^2-2AC)+(A^2-2B)(\xi\chi)_\alpha+(\xi\chi)_\alpha^2].  \notag
\end{align}
This is also a polynomial in the matter potential of maximum fourth order.
In Appendix  \ref{sec:evalUUU} we give this expression which is straightforward to evaluate for $\alpha= e, ~\mu $ and $\tau$. Again, no $\cos\{\frac13\cos^{-1}[\cdots]\}$ terms appear.

For $\alpha=e$ for eq.~\ref{eq:master}, one finds that the right hand side is independent of the matter potential, i.e.
\begin{multline}
\left(\prod_{j>k}\Delta\wh{m^2}_{jk} \right)^2  \left( \prod_i |\wh U_{e i}|^2 \right) =\\
\left(\prod_{j>k}\Delta {m^2}_{jk} \right)^2 \   \left( \prod_i | U_{e i}|^2 \right),
\label{eq:emaster}
 \end{multline}
this is the well known Naumov-Harrison-Scott (NHS) identity \cite{Naumov:1991ju,Harrison:1999df}, divided by the Toshev Identity  \cite{Toshev:1991ku}, squared. In this form, this invariance is not a surprise as in the $a \rightarrow + \infty$ limit 
 \begin{align} 
 \left(\prod_{j>k}\Delta\wh{m^2}_{jk} \right)^2 &    \rightarrow a^4  \, ,  \notag   \\[2mm]
 |\wh U_{e i}|^2 \rightarrow \frac{1}{a^2}    \quad  & \left\{ \begin{array}{l} i= 1, 2  \quad \rm{for ~NO} \\   i=1,3 \quad \rm{for ~IO}  \end{array} 
 \right.  \notag 
 \end{align}
 This implies that the fourth order polynomial must just be a constant.

For $\alpha = \mu$ or $\tau$, again in the   $a \rightarrow +\infty$ limit 
\begin{align} 
 |\wh U_{\mu  i}|^2 \rightarrow \frac{1}{a^2}    \quad   & \left\{ \begin{array}{l} i= 3  \quad \rm{for ~NO} \\   i=2 \quad \rm{for ~IO}  \end{array} 
 \right.  \notag 
 \end{align}
This implies that
 RHS of eq.~\ref{eq:master} is a quadratic polynomial in the matter potential, $a$.  This fact is confirmed by explicit calculation where the coefficients  of the powers the matter potential, $a$, depend on $\theta_{23}$ and $\cos \delta$.   
Further details are also given in  Appendix \ref{sec:evalUUU}.  



\section{Exact Jarlskog Invariant in Matter} 
\label{sec:exact invariant}

We start  from the exact Naumov-Harrison-Scott (NHS) identity \cite{Naumov:1991ju,Harrison:1999df},  which is that the Jarlskog factor in matter times the product of the $\Delta m^2$ matter is an exact invariant\footnote{The simplest way to ``derive'' this expression is that in the $L/E \rightarrow 0$, eq.~\ref{eq:CPvac} and \ref{eq:CPmat} must be equal.} :
\begin{equation}
\frac{J}{\wh J} =\frac{\Delta\wh{m^2}_{32}\Delta\wh{m^2}_{31}\Delta\wh{m^2}_{21}}{\Delta m^2_{32}\Delta m^2_{31}\Delta m^2_{21}}\,.
\label{eq:NHS}
\end{equation}
While the exact eigenvalues have a very complicated analytic form \cite{Zaglauer:1988gz} due to the presence of the $\cos\{\frac13\cos^{-1}[\cdots]\}$ terms, $J^2/\wh J^2$ can be written as a simple fourth order polynomial of the vacuum parameters and the matter potential, because of the FTSP.

Combining eq.~\ref{eq:ABC} and \ref{eq:Dprod} one obtains the exact expression for  $\left(\prod_{i>j}\Delta\wh{m^2}_{ij}\right)^2$ as a fourth order polynomial in the matter potential, $a$.
This guarantees its factorization in two quadratics as shown by Lodovico de Ferrari in 1540.
We start by writing eq.~\ref{eq:NHS} as follows:
\begin{align}
\frac{J^2}{\wh J^2} &= 1+\sum^{4}_{n=1}  f_n \left( \frac{a}{\Delta m^2_{21}}\right)^n
\label{eq:fourth}
\end{align}
where the $f_n$'s are dimensionless functions of the vacuum oscillation parameters which can be easily  derived from eq.~\ref{eq:Dprod}.
Explicit expressions for the $f_n$'s are given in  Appendix \ref{sec:fns}.
$f_1$ and  $f_2$ are of order  1, where as $f_3$ is of order $\epsilon$ and $f_4$ is of order $\epsilon^2$, where
\begin{equation}
\epsilon \equiv \Delta  m^2_{21}/\Delta m^2_{ee} \sim 0.03\, , 
\end{equation}
and $\Dmsqee \equiv c_{12}^2\Delta m^2_{31}+s_{12}^2\Delta m^2_{31}$, \cite{Nunokawa:2005nx}.
Therefore,  there  is a distinct  hierarchy in the  $f_n$'s.

Eq.~\ref{eq:fourth} can be  exactly factorized as
\begin{align}
\left( \frac{J}{\widehat{J}} \right)^2 &  =  \left(1- 2S^{ex}_{sol} \left( \frac{a}{\Delta m^2_{21}}\right) +T^{ex}_{sol} \left( \frac{a}{\Delta m^2_{21}}\right)^2 \right)    \nonumber \\
 \times  & ~ \left(1- 2S^{ex}_{atm}  \left( \frac{a}{\Delta m^2_{ee}}\right) +T^{ex}_{atm} \left( \frac{a}{\Delta m^2_{ee}}\right)^2  \right)
 \label{eq:exfac}
\end{align}
with
\begin{align}
S^{ex}_{sol}   & =  \frac{1}{4}\left(-f_1+\sqrt{f_1^2+4(y-f_2)}\right)    \label{eq:solcoefS}   \\
T^{ex}_{sol} &= \frac{1}{2} \left(y+\sqrt{y^2-4f_4}  \right)  \label{eq:solcoef}  \\
S^{ex}_{atm}   & =  \frac{1}{4}\left(-f_1-\sqrt{f_1^2+4(y-f_2)}\right)/\epsilon   \label{eq:atmcoefS}   \\
T^{ex}_{atm} &= \frac{1}{2} \left(y-\sqrt{y^2-4f_4}  \right)/\epsilon^2  \label{eq:atmcoef} \, .
\end{align}
The $1/\epsilon$ and  $1/\epsilon^2$ terms for $S^{ex}_{atm}$ and $T^{ex}_{atm}$, respectively, are need to change the $(a/\Delta m^2_{21})$ to $(a/\Delta m^2_{ee})$ in second bracket of eq.~\ref{eq:exfac}.
Except for the $a^3$ term, the coefficients for all other powers of $a$ are satisfied for any value of $ y$.  
To get the correct coefficient for the $ a^3$ terms, $y$ must satisfy the following cubic  equation:
\begin{align}
& y^3+g_2y^2+g_1 y  +g_0 =0 
\label{eq:cubic} \\
{\rm with} & \quad  g_2=-f_2, \quad g_1= f_1 f_3-4f_4   \nonumber \\  
{\rm and}  & \quad g_0=(4f_2-f^2_1) f_4 -f^2_3.\nonumber
\end{align}
The relevant, exact solution to this cubic equation is  
\begin{align}
y &= 2 \sqrt{Q} \cos\left\{\frac{1}{3}\arccos\left[R/\sqrt{Q^3} \right] \right\} -g_2/3
\label{eq:cubicsoln} \\
Q&=(g_2^2-3g_1)/9, \quad R=(9g_1g_2-27g_0-2g^3_2)/54 \, ,  \nonumber
\end{align}
where $y  \approx  f_2 \sim 1$.  The other solutions give  complex coefficients in the factorization of eq.~\ref{eq:exfac}.

Unfortunately  the solution  to  the cubic equation does not provide any analytic insight even though it is the exact solution.  However, because the $f_n$'s have a distinct hierarchy in $\Delta  m^2_{21}/\Delta m^2_{ee} $, so do the $g_n$'s:
$g_2$ is of order 1, whereas $g_1$ and $g_0$ are first and second order in  $\Delta  m^2_{21}/\Delta m^2_{ee} $, respectively. This allows an iterative solution to eq.~\ref{eq:cubic},
\begin{align}
y^{(n)} &= -g_2 -g_1/y^{(n-1)}-g_0/(y^{(n-2)})^2  
\end{align}
with $y^{(0)} =-g_2 $ and $y^{(1)} =-(g^2_2 -g_1)/g_2$.
Therefore,  $y$ can easily be calculated to the required order in  $\Delta  m^2_{21}/\Delta m^2_{ee} $.

Then to obtain the factorization coefficients, $S_{sol}$ thru $T_{atm}$, eqs.~\ref{eq:solcoefS} - \ref{eq:atmcoef}, we use the fact that for neutrino oscillations
 there are two small parameters, 
\begin{equation}
\sin^2  \theta_{13}  \sim 0.02  \quad {\rm and}  \quad  \epsilon \equiv \Delta m^2_{21}/\Delta m^2_{ee} \sim 0.03
\end{equation}
so that one can easily perform a Taylor Series expansion in  these two small qualities to obtain
\begin{align}
S_{sol} & \approx c^2_{13}\cos2\theta_{12} - (s^2_{13} \epsilon)  -  \cos2\theta_{12}(s^2_{13} \epsilon^2) +\cdots   \label{eq:Ssol3}   \\[3mm]
T_{sol} &\approx  c^4_{13} -2 \cos 2 \theta_{12}(s^2_{13} \epsilon) +2 \cos2 \theta_{12} (s^4_{13} \epsilon) \nonumber  \\
& \quad  -2 (1-6s^2_{12}c^2_{12})(s^2_{13}\epsilon^2) +   \cdots   \\[5mm]
S_{atm} & \approx  \cos 2 \theta_{13}  + s^2_{12}c^2_{12}(\epsilon^2) -4s^2_{12}c^2_{12}(s^2_{13}  \epsilon^2) \nonumber \\
& \quad +s^2_{12}c^2_{12} \cos2\theta_{12} ( \epsilon^3  )  +\cdots     \\[3mm]
T_{atm} &\approx 1 + 2 s^2_{12} c^2_{12}( \epsilon^2)+ 2 s^2_{12} c^2_{12} \cos 2 \theta_{12}( \epsilon^3)  +\cdots   \label{eq:Patm3}
\end{align}
The first term in  each  of these equations gives  the  approximation discussed in  DP and the correction to the first term are of ${\cal O}(\epsilon^2)$ or ${\cal O}(s^2_{13} \epsilon)$ as expected.

\begin{figure}[t]
\centering
\vspace{10mm}
\includegraphics[width=0.95\columnwidth]{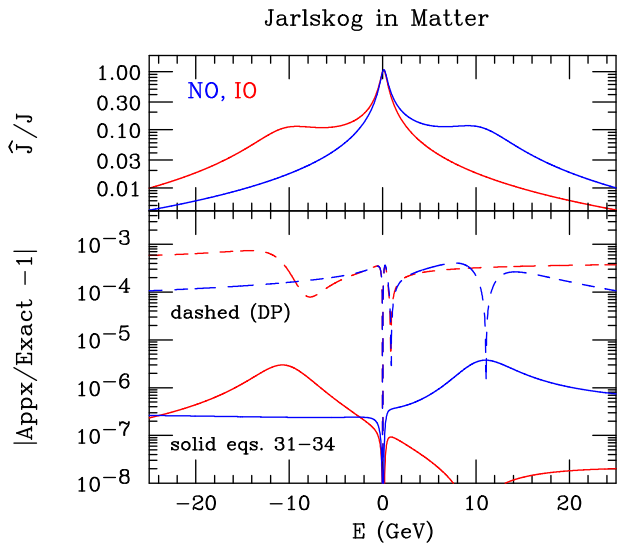}
\vspace{2mm}
\caption{The top panel is the ratio between the Jarlskog invariant in matter and the vacuum Jarlskog invariant for both NO (blue) and IO (red) as function of the energy of the neutrino. Here $Y_e \rho$=1.5 g.cm$^{-3}$. The bottom panel is the fractional difference between the various approximation to the factorization of the Jarlskog invariant compared with the exact Jarlskog invariant in matter. The dashed lines are using the DP approximation, eq.~\ref{eq:Jmat} \cite{Denton:2019yiw} , whereas the solid lines are the third order approximation given in this paper, eqs.~\ref{eq:Ssol3} - \ref{eq:Patm3}, giving more than two orders of magnitude improvement. 
 The results for the exact factorization, eqs.~\ref{eq:solcoefS} - \ref{eq:atmcoef}, are not shown on this figure because they depend on the machine precision; using ``double precision'' one obtains $10^{-14}$ fractional precision. Positive (negative) energy is for neutrinos (anti-neutrinos) and zero is vacuum for both neutrinos and antineutrinos and all energies.}
\label{fig:jarls}
\end{figure}

In Table I we numerical evaluate  the  exact factorization along  with those  of  DP and the approximation  given by eq.~\ref{eq:Ssol3}    to \ref{eq:Patm3}.  

\begin{table*}[t]
\begin{tabular}{||l||c|c||c|c||}
\hline \hline
  ~NO~(IO)        & $~~S_{sol}/S^{ex}_{sol}-1$~~ & ~~$T_{sol}/T^{ex}_{sol}-1$ ~~& ~~$S_{atm}/S^{ex}_{atm}-1$~~ &  ~~$T_{atm}/ T^{ex}_{atm}-|~~$   \\
            \hline
 ~DP approx                                                    & ~~$+1.9~ (-1.9) \times 10^{-3}$ ~~ & ~~$+4.7 ~(-5.0) \times 10^{-4}$   ~~
 &  ~~$-1.9 ~(-1.9) \times 10^{-4}$ ~~& ~~$-4.0 ~ (-4.0) \times 10^{-4}$  ~~  \\
 \hline ~eq.~\ref{eq:Ssol3}  to  \ref{eq:Patm3}  ~~& $-1.6 ~(+1.4) \times 10^{-7}$ &  $-6.1~ (+2.5) \times 10^{-7}$    
 & $+2.8~ (+2.8) \times 10^{-7}$     &  $-6.9 ~(-6.9)  \times 10^{-8}$           \\
 \hline
 \hline
 \end{tabular}
 \caption{The fractional differences for the coefficients of the factorizing quadratics between DP, \cite{Denton:2019yiw}, and the exact, eqs.~\ref{eq:solcoefS} - \ref{eq:atmcoef}, (middle row),  as well as between the perturbative  expansions given in  this paper eqs.~\ref{eq:Ssol3} - \ref{eq:Patm3} and the exact (bottom row).  For this Table the  following  parameters  where used: $\sin^2  \theta_{13}=0.022$,  $\sin^2 \theta_{12}=0.32$ and  $\Delta m^2_{21}= 7.5\times 10^{-5}$ eV$^2$. $\Delta m^2_{ee}= \pm 2.5\times 10^{-3}$ eV$^2$ where the plus (minus) sign is for NO (IO). The matter density times electron fraction used is $Y_e \rho=1.5$ g/cm$^3$. }
 \label{tab:1}
 \end{table*}

Thus the  factorization of the Jarlskog invariant in matter is given by
\begin{equation}
J\approx\Ssol\, \Satm\,\wh{J} \,,
\end{equation}
where
\begin{align}
\Ssol &=\sqrt{1- 2S_{sol} \left( \frac{a}{\Delta m^2_{21}}\right) +T_{sol} \left( \frac{a}{\Delta m^2_{21}}\right)^2  }\,,
\label{eq:Ssol} \\
\Satm &=\sqrt{ 1- 2S_{atm}  \left( \frac{a}{\Delta m^2_{ee}}\right) +T_{atm} \left( \frac{a}{\Delta m^2_{ee}}\right)^2 }\,.
\label{eq:Satm}
\end{align}
This is the same functional form as in DP, eq.~\ref{eq:Jmat}.
Depending  on which approximation one  uses for the $S$'s and  $T$'s one obtains a fractional precision of  $7 \times 10^{-4}$ using DP, eq.~\ref{eq:S13,S12},
$4 \times 10^{-6}$ using  eq.~\ref{eq:Ssol3}-\ref{eq:Patm3} and machine precision using the numerical solution to the cubic equation and eqs.  \ref{eq:solcoefS} - \ref{eq:atmcoef}.
See Fig. \ref{fig:jarls} for both the ratio of $\wh{J} /J$, as well as the fractional difference between the exact and the approximation of  DP,  eq.~\ref{eq:S13,S12} and that of  eq.~\ref{eq:Ssol3}-\ref{eq:Patm3}.  If by machine precision one means ``double precision,'' then 
the fractional difference is $\sim 10^{-14}$.
For phenomenological purposes the precision using the simple result of DP \cite{Denton:2019yiw} is accurate enough for most applications.


\section{Relationship with the Solar and  Atmospheric Resonances}
Its easily shown that $\Ssol $ is minimum when\footnote{For the numerical calculations in this section, we use the same parameters as  in Table 1.}
\begin{align}
 \frac{a^\text{sol}_\text{min}}{ \Delta m^2_{21} } & = \frac{S^{ex}_{sol}}{T^{ex}_{sol}} =  \frac{\cos 2\theta_{12}}{c^2_{13} } 
+  \cos 4 \theta_{12} \,(s^2_{13} \epsilon)  + \cdots    \label{eq:asolx}  \\[2mm]
 & =  \frac{\cos 2\theta_{12}}{c^2_{13} } 
+\left\{ \begin{array}{ll}
-5.2 \times 10^{-4}  \quad & \text{NO} \\
~~5.0 \times 10^{-4}  \quad & \text{IO} 
\end{array} \right.
 \, ,  \notag
\end{align}
and  that $\Satm$ is minimum when
\begin{align}
 \frac{a^\text{atm}_\text{min}}{ \Delta m^2_{ee} }  &  = \frac{S^{ex}_{atm}}{T^{ex}_{atm}} = \cos 2\theta_{13} -s^2_{12} c^2_{12}( \epsilon ^2 )
+ \cdots   \label{eq:aatmx}  \\[2mm]
& =  \cos 2\theta_{13} 
- \left\{ \begin{array}{ll}
2.0 \times 10^{-4}  \quad & \text{NO} \\
1.9 \times 10^{-4}  \quad & \text{IO} 
\end{array} \right.
\, .\notag
\end{align}
from  eq.~\ref{eq:Ssol} and \ref{eq:Satm} respectively. The ``$\cdots$'' are higher order terms in $s^2_{13}$ and $\epsilon$. Clearly, these two values for the matter potential are close to the values for the solar and atmospheric resonances.  But what is the precise relationship? It cannot be exact as both  $\Ssol $  and  $\Satm $ are symmetric functions about the value that minimizes them, whereas for the solar and atmospheric resonances there must be some asymmetry caused by the fact that the solar resonance is below (above) atmospheric resonance for the NO (IO).

To answer this question we first have to define the solar and atmospheric resonance.  For the solar resonance we define it to be the value of the matter potential that  minimizes the separation between the matter mass eigenstates $|\wh{\nu}_1\rangle$ and $ |\wh{\nu}_2 \rangle $, that is when
\begin{align}
\frac{d  (\wh{m^2}_{2}-  \wh{m^2}_{1} ) }{da} = 0 \, . 
\label{eq:solres}
\end{align}
It can be easily shown by taking the derivative of A, B \& C with respect to $a$, see Appendix \ref{sec:dmsqda}, or using the evolution equations of  \cite{Xing:2018lob} that
\begin{align}
\frac{d \wh{m^2}_{i}}{  da }  =  |\wh{U}_{ei}|^2\, .
\label{eq:deriv}
\end{align}
Therefore the solar resonance condition also implies that
\begin{align}
  |\wh{U}_{e1}|^2=|\wh{U}_{e2}|^2,  \quad  \sin^2 \wh{\theta}_{12}=0.5\,.
 \end{align}
 That is, by minimizing the separation of the matter mass eigenstates $|\wh{\nu}_1\rangle$ and $ |\wh{\nu}_2 \rangle $, we also have {\it exact}  maximal mixing  between the matter eigenstates $|\wh{\nu}_1\rangle$ and $ |\wh{\nu}_2 \rangle $.

For the atmospheric resonance, one could consider the resonance condition for NO to be 
\begin{align}
\frac{d \, (\wh{m^2}_{3} - \wh{m^2}_{2} )}{da} = 0 \, ,   \quad   |\wh{U}_{e3}|^2=|\wh{U}_{e2}|^2 \, ,
\label{eq:32res}
\end{align}
which implies $ \sin^2 \wh{\theta}_{13} \approx 0.5.$
 For IO
\begin{align}
\frac{d \, (\wh{m^2}_{3} - \wh{m^2}_{1} )}{da} = 0 \, , \quad  |\wh{U}_{e3}|^2=|\wh{U}_{e1}|^2 \, ,
\label{eq:31res}
\end{align}
which also implies $ \sin^2 \wh{\theta}_{13} \approx 0.5$, \cite{Xing:2018lob}. That is, {\it almost} maximal mixing between  $|\wh{\nu}_3\rangle$ and  $|\wh{\nu}_2\rangle$  (\,$|\wh{\nu}_1\rangle$\,) for  NO  (IO).    For both NO and IO,  the fractional difference between $\sin^2 \wh{\theta}_{13}$ and  0.5  is
 $ \sim 2 \times 10^{-4}$ with these definitions.

However, we present here a mass ordering independent definition of the atmospheric resonance, given by
\begin{align}
\frac{d \, (\wh{m^2}_{3} - \wh{m^2}_{2} -  \wh{m^2}_{1} )}{da} = 0 \, . 
\label{eq:atmres}
\end{align}
This gives an excellent approximation\footnote{The fractional difference between the solution to  eqs.~\ref{eq:32res} and \ref{eq:atmres} or   eqs.~\ref{eq:31res} and \ref{eq:atmres} is $7\times 10^{-5}$. Using eqs.~\ref{eq:32res}  and  \ref{eq:31res} as the definition of the atmospheric resonance does not change our conclusions. } to eq.~\ref{eq:32res} for NO  as the mass of the matter eigenstate $|\wh{\nu}_1\rangle$ is essentially independent of the matter potential at the atmospheric resonance as $  d \wh{m^2}_{1} / da = |\wh{U}_{e1}|^2 \approx 2 \times 10^{-4}$ here. Similarly for  eq.~\ref{eq:31res} for IO.

This mass ordering independent definition also gives
\begin{align}
  |\wh{U}_{e3}|^2=|\wh{U}_{e2}|^2+|\wh{U}_{e1}|^2,  \quad  \sin^2 \wh{\theta}_{13}=0.5\,,
 \end{align}
 from eq.~\ref{eq:deriv}, independent of the mass ordering.
 That is, {\it exact} maximal mixing between the matter eigenstate $|\wh{\nu}_3\rangle$ and the state $ \cos \wh{\theta}_{12} |\wh{\nu}_1 \rangle + \sin \wh{\theta}_{12} |\wh{\nu}_2 \rangle $. Therefore, the mass ordering independent definition, eq.~\ref{eq:atmres},  is what will be used for the atmospheric resonance in the rest of this paper.

Independent of this discussion, it was argued in \cite{Denton:2018cpu} that  combination $(\wh{m^2}_{3} - \wh{m^2}_{2} -  \wh{m^2}_{1})$ is the effective $\Delta m^2_{ee} (a)$  in matter with the addition of a constant that depends on ones conventions:
$\Delta m^2_{ee} (a)$ gives the effective frequency for a $\nu_e$ disappearance oscillations in matter at the atmospheric minima and coincides with $\Delta m^2_{ee}$ in vacuum, i.e. $\Delta m^2_{ee} (a=0)=\Delta m^2_{ee}$.  
 In the conventions of this paper,
\begin{align}
& \Delta m^2_{ee} (a)  \equiv \wh{m^2}_{3} - \wh{m^2}_{2} -  \wh{m^2}_{1} + \Delta m^2_{21} c^2_{12} \\
& \approx \Delta m^2_{ee} \sqrt{  (1- 2\cos 2 \theta_{13}(a/ \Delta m^2_{ee})+ (a/\Delta m^2_{ee})^2 }\,. \notag
\end{align}
Note, this approximation is just the leading terms in $\Delta m^2_{ee}\Satm$. \\

With the above definitions, the exact position of the solar and atmospheric resonances, $a_{\text{xsol}}$ and $a_{\text{xatm}}$ respectively, can be calculated perturbatively, see Appendix \ref{sec:pert}, and numerical from eq.~\ref{eq:whmsq} and we find that 
\begin{align}
\frac{a_{\text{xsol}} }{ \Delta m^2_{21} }  & =  \frac{\cos 2\theta_{12}}{c^2_{13} } 
+ \cos^2 2\theta_{12} (s^2_{13} \epsilon)
 +\cdots \label{eq:axsol}  \\
& =  \frac{\cos 2\theta_{12}}{c^2_{13} } 
+ \left\{ \begin{array}{ll}
~~9.7 \times 10^{-5}  \quad & \text{NO} \\
-8.2 \times 10^{-5}  \quad & \text{IO} 
\end{array} \right.
 \, ,
\notag
\end{align}
and
\begin{align}
 \frac{a_{\text{xatm}}}{ \Delta m^2_{ee} }  & = \cos 2\theta_{13} 
- s^2_{12} c^2_{12}(1-2s_{13}) \, ( \epsilon^2 )+\cdots
\,   \label{eq:axatm} \\
& = \cos 2\theta_{13} 
- \left\{ \begin{array}{ll}
1.52\times 10^{-4}  \quad & \text{NO} \\
1.49\times 10^{-4}  \quad & \text{IO} 
\end{array} \right.
\, .
\notag
\end{align}
Thus the accuracy of the leading terms is at the $10^{-4}$ level, provide one includes the $c^2_{13}$ 
for the solar resonance and $\Delta m^2_{ee}$ instead of $\Delta m^2_{31}$ or $\Delta m^2_{32}$ for the atmospheric resonance,  \cite{Zaglauer:1988gz, Denton:2016wmg}.  The magnitude of $\Delta m^2_{31}$ and $\Delta m^2_{32}$ differs from  $\Delta m^2_{ee}$ at the 1-2\% level. 

\begin{figure}[t]
\centering
\includegraphics[width=0.8\columnwidth]{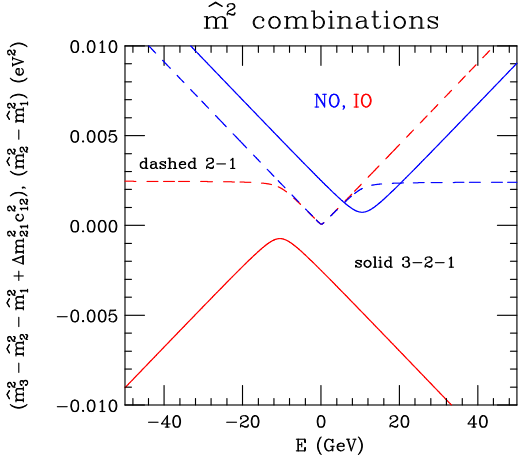}
\caption{The solid lines are $(\wh{m^2}_3-\wh{m^2}_2-\wh{m^2}_1 +\Delta m^2_{21} c^2_{12})$ for both NO (blue) and IO (red) as a function of the neutrino energy. Note, the approximate symmetry about the minima, $|E|\sim$ 10.5 GeV, for these two functions. At $E=0$, i.e. vacuum, this combination is exactly $\pm \Delta m^2_{ee}$, plus NO, minus for IO. The dashed lines are $(\wh{m^2}_2-\wh{m^2}_1)$  for NO (blue) and IO (red). These two functions have a significant asymmetry about the minima, E $\sim$ 0.12 GeV, especially for $|E|>$ 10 GeV.  Here, $Y_e \rho = 1.5$ g.cm$^{-3}$.
Positive (negative) energy is for neutrinos (anti-neutrinos) and zero is vacuum for both neutrinos and antineutrinos and all energies.}
\label{fig:msqs}
\end{figure}

Comparing the eqs.~\ref{eq:asolx} \& \ref{eq:axsol}  and eqs.~\ref{eq:aatmx} \& \ref{eq:axatm}, it clear that the minima of $\Ssol$ and $\Satm$
do not exactly coincide with the exact solar and exact atmospheric resonances but the difference is at the 0.06\% for solar and 0.005\% for atmospheric which are negligible for phenomenological purposes.  The order of magnitude difference between these two is probably due to the fact that the variable $(\wh{m^2}_{3} - \wh{m^2}_{2} -  \wh{m^2}_{1})$ is more the symmetric about the atmosphere resonance than $(\wh{m^2}_{2} -  \wh{m^2}_{1})$ is about the solar resonance, see Fig. \ref{fig:msqs}.

It is also worth noting that by  using  $ \Delta m^2_{ee} |_{IO} = -\Delta m^2_{ee}|_{NO}$  rule for switching between the two mass orderings (NO $\leftrightarrow$ IO), the fractional difference between  $|a_\text{xatm}|$ for the two orderings is $\sim 3\times 10^{-6}$ 
and in fact, the fractional difference between
\begin{multline}
-(\wh{m^2}_{3} - \wh{m^2}_{2} - \wh{m^2}_{1} + \Delta m^2_{21} c^2_{12}  )|_{IO}(-a)\\ \quad \text{and} \quad  (\wh{m^2}_{3} - \wh{m^2}_{2} -  \wh{m^2}_{1} + \Delta m^2_{21} c^2_{12}  )|_{NO}(a)
\end{multline}
 is smaller than $10^{-5}$ for all $a$.
As compared to $\sim $1\%,  if one uses  any of the following rules to flip the mass ordering:
\begin{multline}
\Delta m^2_{31} |_{IO} = -\Delta m^2_{31}|_{NO} \quad \text{or}  \quad \Delta m^2_{32} |_{IO} = -\Delta m^2_{32}|_{NO}\\
  \text{or}  \Delta m^2_{32} |_{IO} = -\Delta m^2_{31}|_{NO} \, . \quad \quad  \notag
 \end{multline}

 

\section{Summary and Conclusions}
\label{sec:conclusions}

In this paper we have derived three new results for three flavor neutrino oscillations in matter.
First, we have shown that any quantity that can be represented by a symmetric polynomial of the eigenvalues of the Hamiltonian, 
$\wh{m^2}_j$, can be written as a polynomial in the matter potential, $a$.  Examples of such quantities are
\begin{align}
 &  \left(\prod_{j>k}\Delta\wh{m^2}_{jk} \right)^2, \quad  \left( \frac{J}{\widehat{J}} \right)^2 \,  
 \notag \\
 \text{and} \quad &  \left(\prod_{j>k}\Delta\wh{m^2}_{jk} \right)^2 \times \left( \prod_i |\wh U_{\alpha i}|^2 \right) \, . \notag
\end{align}
The first two are fourth order polynomials in the matter potential whereas the last one is a constant for $\alpha =e$ and a quadratic polynomial in the matter potential for $\alpha=\mu, ~ \tau.$ Clearly there are many other such quantities that are symmetric in the eigenvalues.

Second, for the Jarlskog invariant in matter we give the exact factorization of the fourth order polynomial, such that
\begin{equation}
\wh{J} = \frac{J}{\Ssol^\text{ex}\, \Satm^\text{ex}}\,, \notag
\end{equation}where
\begin{align}
\Ssol^{ex} &=\sqrt{1- 2S^{ex}_{sol} \left( \frac{a}{\Delta m^2_{21}}\right) +T^{ex}_{sol} \left( \frac{a}{\Delta m^2_{21}}\right)^2  }\,, \notag \\
\Satm^{ex} &=\sqrt{ 1- 2S^{ex}_{atm}  \left( \frac{a}{\Delta m^2_{ee}}\right) +T^{ex}_{atm} \left( \frac{a}{\Delta m^2_{ee}}\right)^2 }\,. \notag
\end{align}
The exact coefficients $S^{ex}_{sol}$ to $T^{ex}_{atm}$ are given in eqs.~\ref{eq:solcoefS} - \ref{eq:atmcoef}. Since the exact coefficients involve the solution to a cubic polynomial which are notoriously challenging to penetrate analytically,
a Taylor series expansion of these  coefficients to
${\cal O}(s^4_{13}\epsilon ), ~ {\cal O}(s^2_{13}\epsilon^2 ),~{\cal O}(\epsilon^3 ),$
is also given in eqs.~\ref{eq:Ssol3}  --  \ref{eq:Patm3}.
The fractional accuracy of these coefficients is better than $10^{6}$, as shown in Table~1.

Third,  we define the solar resonance to be when
\begin{align}
\frac{d  (\wh{m^2}_{2}-  \wh{m^2}_{1} ) }{da} = 0  \quad \Longleftrightarrow \quad   \sin^2 \wh{\theta}_{12}=0.5 \, , \notag
\end{align}
and calculate that this occurs when
\begin{align}
\frac{a }{ \Delta m^2_{21} }  & =  \frac{\cos 2\theta_{12}}{c^2_{13} } 
+ \cos^2 2\theta_{12} (s^2_{13} \epsilon)
 +\cdots \,. \notag
 \end{align}
 where the ``$\cdots$'' are higher order terms in $s^2_{13}$ and $\epsilon \equiv \Delta m^2_{21}/\Delta m^2_{ee}$.
Independent of the mass ordering, we define the atmospheric resonance to be when
\begin{align}
\frac{d  (\wh{m^2}_{3}-  \wh{m^2}_{2}- \wh{m^2}_{1} ) }{da} = 0  \quad \Longleftrightarrow \quad   \sin^2 \wh{\theta}_{13}=0.5 \, , \notag
\end{align}
and calculate that this occurs when
\begin{align}
 \frac{a}{ \Delta m^2_{ee} }  & = \cos 2\theta_{13} 
- s^2_{12} c^2_{12}(1-2s_{13}) \, ( \epsilon^2 )+\cdots
\, . \notag 
\end{align}
For three flavors of neutrinos, these are the most accurate values for the locations of the solar and atmospheric resonances available in the literature.
The value of the matter potential typically quoted for the atmospheric resonance, $a/\Delta m^2_{32}=\cos 2 \theta_{13}$, differs from above value by 2\%, a phenomenological significant difference.

We have also made a comparison between the location of the minima in the matter potential for the two factorizing factors, $\Ssol^{ex} $ and $\Satm^{ex} $ and the location in matter potential of the solar and atmospheric resonances.
The difference in the value of the matter potential between the location of the minima and the resonance values are  0.06\% for solar and 0.005\% for atmospheric, therefore they do not coincide.  However, these difference are negligible for phenomenological purposes.  

To conclude, the understanding of the physics of neutrino propagation in matter is of essential importance to the current NOvA and T2K,  and upcoming DUNE and T2HK(K) long baseline neutrino oscillation experiments. The Wolfenstein matter effect significantly complicates three flavor neutrino oscillations in matter compared to vacuum. The exact analytic expressions, even for constant matter density, are analytically impenetrable due to the presence of the $\cos\{\frac13\cos^{-1}[\cdots]\}$ terms. In this paper we show that variables that depend on the symmetric polynomials of the mass squared of the neutrino eigenstates in matter are just simple polynomials of the matter potential, a significant simplification.  We give a number of examples and in particular the Jarlskog invariant in matter which controls the size of intrinsic CP violation in the above neutrino appearance experiments. We give for the first time the exact factorization of the Jarlskog invariant in matter into two quadratic resonance factors. The location of the minima of these resonance factors is compared to the location of the solar and atmospheric resonances to the appropriate order in the small quantities   $s^2_{13}$ and $\Delta m^2_{21}/\Delta m^2_{ee}$.  This required, for the first time, the calculation of the location of the solar and atmospheric resonances beyond leading order.   All of this further enhances our understanding of neutrino propagation in matter relevant for the long baseline neutrino oscillation experiments being currently performed and for the next generation of experiments.
\\  

\newpage

\begin{widetext}

\begin{acknowledgments}
I thank Peter Denton for many enlightening discussions during the course of this work.
Fermilab is operated by the Fermi Research Alliance under contract no.~DE-AC02-07CH11359 with the U.S.~Department of Energy. 
This project has received funding/support from the European Union's Horizon 2020 research and innovation programme under the Marie 
Sklodowska-Curie grant agreement No 860881-HIDDeN.

I dedicate this paper to the memory of the outstanding neutrino theorist and friend, \href{https://inspirehep.net/literature?sort=mostrecent&size=25&page=1&q=a\%20S.Pakvasa.1\%20and\%20ac\%201-\%3E9}{Sandip Pakvasa}.  His insights on neutrino physics and the neutrinos he emitted, will live on until the end of the Universe.
\end{acknowledgments}

\appendix

\section{The variables $\xi$ and $\chi$}
\label{sec:xichi}
The quantities $(\xi+\chi)_\alpha$ and $(\xi\chi)_\alpha$ are the trace and determinant of  the sub-matrix of (2E)H with the $\alpha$-row and $\alpha$-column removed.
In the flavor basis, since $H_{ee}$ is the only element that depends on the matter potential these traces and determinants are at most linear polynomials in the matter potential, a. In fact,  $(\xi+\chi)_e$ and $(\xi\chi)_e$ are independent of the matter potential.  Explicitly,

\begin{align}
\xi_e+\chi_e&=
\left(\h_{\mu\mu}+\h_{\tau\tau}\right)  =\Delta m^2_{31}c^2_{13}+\Delta m^2_{21}(c^2_{12}+s^2_{13}s^2_{12} )= d B/da
\,,\label{eq:sub e sum}\\
\xi_e\chi_e&=
\left(\h_{\mu\mu}\h_{\tau\tau}-\h_{\mu\tau}^2\right)  = \Delta m^2_{31} \Delta m^2_{21} c^2_{13} c^2_{12}= d C/da
\,,\label{eq:sub e prod}\\[2mm]
\xi_\mu+\chi_\mu&=
\left[c_{23}^2(\h_{ee}+\h_{\tau\tau})+s_{23}^2(\h_{ee}+\h_{\mu\mu})-2s_{23}c_{23}\cos\delta (\h_{\mu\tau})  \right]\,,\label{eq:sub mu sum}\\
\xi_\mu\chi_\mu&=
\left[ c_{23}^2(\h_{ee}\h_{\tau\tau}-\h_{e\tau}^2)+s_{23}^2(\h_{ee}\h_{\mu\mu}-\h_{e\mu}^2)
-2s_{23}c_{23}\cos\delta(\h_{ee}\h_{\mu\tau}-\h_{e\mu}\h_{e\tau})\right]    \,.\label{eq:sub mu prod}
\end{align}
The $\xi_\tau$ and $\chi_\tau$ eigenvalues are the same as $\xi_\mu$ and $\chi_\mu$ under the interchange $s_{23}^2\leftrightarrow c_{23}^2$ and $s_{23}c_{23}\to-s_{23}c_{23}$. Where
\begin{equation}
\h=
\begin{pmatrix}
a+\Dmsqee s_{13}^2+\Delta m^2_{21}s_{12}^2&\quad&c_{13}s_{12}c_{12}\Delta m^2_{21}&&s_{13}c_{13}\Dmsqee\\[0.5em]
\cdot&&\Delta m^2_{21}c_{12}^2&\quad&-s_{13}s_{12}c_{12}\Delta m^2_{21}\\[0.5em]
\cdot&&\cdot&&\Dmsqee c_{13}^2+\Delta m^2_{21}s_{12}^2
\end{pmatrix}
\,,
\label{eq:Hrot0}
\end{equation}
where $\h_{\alpha\beta}=\h_{\beta\alpha}$ and $\Dmsqee\equiv c_{12}^2\Delta m^2_{31}+s_{12}^2\Delta m^2_{31}$ \cite{Nunokawa:2005nx}. 
 The relationship between $\h$ and H of eq.~\ref{eq:ham} is\\ $\h=
 (2E)  ~ U^\dagger_{23}(\theta_{23},\delta)H U_{23}(\theta_{23},\delta)$.
See \cite{Denton:2019ovn} for further details.

Defining  $\xi_\alpha \leq  \chi_\alpha$, then  the Cauchy interlace theorem tells us that
$ \wh{m^2}_1 \leq \xi_\alpha \leq  \wh{m^2}_2   \leq  \chi_\alpha \leq \wh{m^2}_3$ for NO.

\section{ Evaluation of $\prod_i \left((\wh{m^2}_i)^2 -(\xi+\chi)_\alpha \wh{m^2}_i+(\xi \chi)_\alpha \right)$}
\label{sec:evalUUU}

It is straight forward to show that 
\begin{align}
& \prod_i \left((\wh{m^2}_i)^2 -(\xi+\chi)_\alpha \wh{m^2}_i+(\xi \chi)_\alpha \right) \notag \\
=  & ~~C[C-B(\xi+\chi)_\alpha+A(\xi+\chi)_\alpha^2-(\xi+\chi)_\alpha^3]  \label{eq:UUU} \\
  & -(\xi+\chi)_\alpha(\xi\chi)_\alpha [(AB-3C)-B(\xi+\chi)_\alpha+A(\xi\chi)_\alpha ]  \notag  \\
  &+(\xi\chi)_\alpha [(B^2-2AC)+(A^2-2B)(\xi\chi)_\alpha+(\xi\chi)_\alpha^2] \,. \notag
\end{align}
For $\alpha =e$,  both $(\xi+\chi)_e$ and $(\xi\chi)_e$ are independent of $a$ and this is at most a quadratic polynomial in $a$. The quadratic and linear terms in $a$ cancel and one is left with the constant, so this expression is independent of the matter potential.  

For $\alpha =\mu$ (or $\tau$),  both $(\xi+\chi)_\mu$ and $(\xi\chi)_\mu$ are linear in $a$ and this polynomial is at most fourth order in $a$.
Again the top two powers of $a$ cancel and one is left with a quadratic polynomial in $a$.  Since  $(\xi+\chi)_\mu$ and $(\xi\chi)_\mu$ depend on $\theta_{23}$ and $\delta$ it is convenient to reorganize the calculation by multiplying $(\wh{m^2}_i)^2$ by $(c^2_{23}+s^2_{23})$ as follows:
\begin{align}
 \prod_i \left((\wh{m^2}_i)^2 -(\xi+\chi)_\mu \wh{m^2}_i+(\xi \chi)_\mu \right)  =  \prod_i \left( c^2_{23} X_i + s^2_{23} Y_i +2s_{23}c_{23}\cos\delta ~Z_i \right)
\end{align}
where
\begin{align}
X_i &=  (\wh{m^2}_i)^2 - (\h_{ee}+\h_{\tau\tau}) (\wh{m^2}_i) +(\h_{ee}\h_{\tau\tau}-\h_{e\tau}^2) =(\wh{m^2}_i-\h_{ee})(\wh{m^2}_i-\h_{\tau \tau})-\h^2_{e\tau} \, , \notag \\
Y_i &=   (\wh{m^2}_i)^2 - (\h_{ee}+\h_{\mu\mu}) (\wh{m^2}_i) +(\h_{ee}\h_{\mu \mu}-\h_{e\mu}^2)=(\wh{m^2}_i-\h_{ee})(\wh{m^2}_i-\h_{\mu \mu})-\h^2_{e\mu}   \, , 
  \\
Z_i &=  \h_{\mu\tau} (\wh{m^2}_i ) -(\h_{ee}\h_{\mu\tau}-\h_{e\mu}\h_{e\tau})  =  \h_{\mu\tau} (\wh{m^2}_i  -\h_{ee})  -\h_{e\mu}\h_{e\tau}  \, , 
\notag
\end{align}
which satisfy the following condition that $X_i Y_i = Z^2_i$. Then the LHS of eq.~\ref{eq:UUU} is a homogeneous third order polynomial in
$c^2_{23}$, $s^2_{23}$ and $(2s_{23}c_{23}\cos\delta)$:
 \begin{align}
 &  \prod_i \left( c^2_{23} X_i + s^2_{23} Y_i +2s_{23}c_{23}\cos\delta ~Z_i \right) \notag \\
 =& ~ c_{23}^6 ~ [X_1X_2X_3]  + c_{23}^4  s^2_{23}  ~[X_1X_2Y_3]  + c_{23}^2   s^4_{23}  ~[X_1Y_2Y_3] + s_{23}^6~[Y_1Y_2Y_3] \notag \\[3mm]
& + \, (2s_{23}c_{23} \cos \delta ) (~c_{23}^4 ~[Z_1X_2X_3]  
 + c_{23}^2s_{23}^2 ~[X_1Y_2Z_3] 
 + s_{23}^4 ~[Z_1Y_2Y_3]  ~ ) \notag \\[3mm]
& + \, (2s_{23}c_{23} \cos \delta)^2  (c_{23}^2~[Z_1Z_2X_3] +s_{23}^2 ~[Z_1Z_2Y_3] ~)   +\, (2s_{23}c_{23} \cos \delta)^3 ~[Z_1Z_2Z_3 ]
\end{align}
where each $[\cdots]$ is symmeterized over the labels (1,2,3) which guarantees that each term is a polynomial in $a$.
This is easy to calculate using an algebraic program like FORM. Some, but not all of the terms are very simple, we give here examples of the simplest terms.  


The coefficient of $s^6_{23}$ is given by
$$ Y_1 Y_2 Y_3 =s^4_{13} c^2_{13} ~s^2_{12} c^2_{12}~ (\Delta m^2_{21} )^2 \biggr(  \Delta m^2_{31} \Delta m^2_{32} +  a \,  \Delta m^2_{ee} \biggr)^2,$$
the coefficient of $c^6_{23}$ is
$$  X_1 X_2 X_3 =s^2_{13} c^2_{13}~  s^4_{12} c^4_{12}~ (\Delta m^2_{21} )^4   \, a^2 , $$
and the coefficient of  $(2s_{23}c_{23}\cos\delta)^3$
$$ Z_1 Z_2 Z_3=  s^3_{13} c^2_{13} ~ s^3_{12} c^3_{12} ~ (\Delta m^2_{21})^3( \Delta m^2_{31} \Delta m^2_{32}+a  \Delta m^2_{ee}  )a \, . $$
Note $Z_1 Z_2 Z_3=\sqrt{ X_1 X_2 X_3 \, Y_1 Y_2 Y_3}$.\\

For $\alpha=\tau$, just interchange $s^2_{23} \leftrightarrow c^2_{23}$ and flip sign of $s_{23} c_{23}$.

\section{Derivation of the $f_n$'s}
\label{sec:fns}

Writing A, B and C, defined in eq.~\ref{eq:ABC} , as constant plus linear in $a$  terms, as follows,
\begin{align}
A&=A_0+a, \quad \quad ~A_0=\Delta m^2_{21}+\Delta m^2_{31}, \notag \\
B&=B_0+aB_1, \quad B_0=\Delta m^2_{31}\Delta m^2_{21}, \quad \quad \quad  B_1=\Delta m^2_{31}c^2_{13}+\Delta m^2_{21}(c^2_{12}+s^2_{13} s^2_{12}), \notag \\
C&=aC_1, \quad \quad \quad ~~C_1=\Delta m^2_{31}\Delta m^2_{21}c^2_{13} c^2_{12}.
\end{align}
Then from eq.~\ref{eq:Dprod}, the $f_n$'s of eq.~\ref{eq:fourth} are given by 
\begin{align}
f_1 &= 2 \Delta m^2_{21} [(A_0^2 - 4B_0)(B_0B_1 - 2A_0C_1) + B_0^2(A_0 - 2B_1) + A_0B_0C_1 \,] /f_0  \, ,  \notag \\
f_2 & = (\Delta m^2_{21})^2\,[(A_0^2 - 4B_0)(B_1^2 - 4C_1) + B_0^2 + 
    4(A_0 - 2B_1)(B_0B_1 - 2A_0C_1) + 2(B_1A_0 + B_0)C_1 - 27C_1^2 \, ]/f_0   \, ,   \notag \\
f_3 &= 2 (\Delta m^2_{21})^3 [(B_0B_1 - 2A_0C_1) + (B_1^2 - 4C_1)(A_0 - 2B_1) +  B_1C_1] /f_0  \, ,  \notag \\
f_4 &=(\Delta m^2_{21})^4 (B_1^2 - 4C_1) /f_0  \, , 
\end{align}
where $f_0 = B_0^2(A_0^2 - 4B_0)=(\Delta m^2_{21}\Delta m^2_{31}\Delta m^2_{32})^2$. \\

Note, $f_1, ~f_2, ~f_3 ~\& ~f_4$ are of order 
1, 1, $\epsilon$ \& $\epsilon^2$, respectively, where $\epsilon \equiv \Delta m^2_{21}/\Delta m^2_{ee}$.

\vspace{5mm}

\section{Proof of $d \,  \wh{m^2}_i /da = |\wh{U}_{ei}|^2$}
\label{sec:dmsqda}

Differentiate A, B \& C of eq.~\ref{eq:ABC} with respect to $a$, to obtain the following:
\begin{align}
1 & = \frac{d \wh{m^2}_1 }{da }+\frac{d \wh{m^2}_2 }{da }+\frac{d \wh{m^2}_2 }{da }\, ,  \label{eq:dA} \\
 (\xi+\chi)_e  &=( \wh{m^2}_2+ \wh{m^2}_3) \frac{d \wh{m^2}_1 }{da }+( \wh{m^2}_3+ \wh{m^2}_1) \frac{d \wh{m^2}_2 }{da }+( \wh{m^2}_1+ \wh{m^2}_2) \frac{d \wh{m^2}_3 }{da }   \, , \label{eq:dB} \\
 (\xi \chi)_e &=( \wh{m^2}_2 \wh{m^2}_3) \frac{d \wh{m^2}_1 }{da }+( \wh{m^2}_3 \wh{m^2}_1) \frac{d \wh{m^2}_2 }{da }+( \wh{m^2}_1 \wh{m^2}_2) \frac{d \wh{m^2}_3 }{da }   \, . \label{eq:dC}
\end{align}
Multiple eq.~\ref{eq:dA} by $ (\wh{m^2}_i)^2$, eq.~\ref{eq:dB} by $ (-\wh{m^2}_i)$ and add these to eq.~\ref{eq:dC}, one obtains
\begin{align}
 \frac{d \, \wh{m^2}_i }{da } & =\frac{(\wh{m^2}_i)^2 -(\xi+\chi)_e (\wh{m^2}_i) +(\xi \chi)_e}{\Delta \wh{m^2}_{ij}\Delta \wh{m^2}_{ik}} = |\wh{U}_{ei}|^2 \, .
 \end{align}
 with the indicies $i$, $j$ and $k$ all different.  See eq.~\ref{eq:Uaisq}.
 
 \section{Perturbative Expansions about the Resonances}
 \label{sec:pert}
 
A perturbative expansion is used to obtain the values of the matter potential for the atmospheric and solar resonances.\\
 
\subsection{Atmospheric Resonance}
In this Appendix we perform an perturbative expansion in the matter potential about the approximate value of the matter potential that minimizes  
$(\wh{m^2}_3 -\wh{m^2}_2 -\wh{m^2}_1)$, that is 
\begin{equation}
a= \Delta m^2_{ee} \cos 2 \theta_{13} +\delta a
\label{eq:aRatm}
\end{equation}
Starting with the Hamiltonian from eq.~\ref{eq:ham}, a (2-3) rotation using $U_{23}(\theta_{23},\delta)$ is performed to remove the $\theta_{23}$ and $\delta$ dependence. Then a $\pi/4$ rotation is performed in the (1-3) sector using the value of $a$ given in eq.~\ref{eq:aRatm}. The resulting Hamiltonian is given by 
\begin{align}
U^\dagger_{13}(\pi/4)U^\dagger _{23}(\theta_{23},\delta)~ (2E)H ~ U_{23}(\theta_{23},\delta)U_{13}(\pi/4)  = H_a+h
\end{align} 
with
\begin{align}
H_a  =\text{\bf diag}(~& \Delta m^2_{ee} (c^2_{13} -s_{13} c_{13})+\Delta m^2_{12} s^2_{12},  ~~ \Delta m ^2_{21} c^2_{12} ,~~\Delta m^2_{ee} (c^2_{13} +s_{13} c_{13})+\Delta m^2_{12} s^2_{12}  ~ ) \, ,  \\[3mm]
& \quad  h = \frac12 \left( 
 \begin{array}{ccc}
\delta a \quad & \delta m^2_+  \quad &  \delta a  \\
\cdot & 0  \quad &   \delta m^2_- \\
\cdot &  \cdot  &  \delta a 
\end{array}  \right)   \label{eq:hatm}
\end{align}
where $\delta m^2_\pm =\sqrt{2} \Delta m^2_{21} s_{12} c_{12}  (c_{13}\pm s_{13}) $. Note, because the solar crossing has not been resolved the mass eigenstates $\nu_1$ and $\nu_2$ are interchanged.  Since $h$ is symmetric, the $\cdots$'s are given by the appropriate term above the diagonal.

Now perturbation theory can be used to calculate $(\wh{m^2}_3 -\wh{m^2}_2 -\wh{m^2}_1)$, which to third order in $h$, is given by 
\begin{align}
&\wh{m^2}_3 -\wh{m^2}_2 -\wh{m^2}_1 \approx ~ d_{31}+2(h^2_{13}/d_{31} + h^2_{23}/d_{32})+ 2h_{11} h_{23} (2 h_{12} d_{32}-h_{23} d_{31})/(d_{32}^2 d_{31})\, ,
\label{eq:321pert}
\end{align} 
where $h_{jk}$ are the elements of matrix $h$, eq.~\ref{eq:hatm}, and  $d_{jk}=H_a(j,j)-H_a(k,k)$, e.g. $d_{31}=\Delta m^2_{ee} (2 s_{13} c_{13})$. Eq.~\ref{eq:321pert} is a quadratic in the variable $\delta a$ as both $h_{11}$ and $h_{13}$ are linear in $\delta a$.  
To obtain an estimate of the shift one needs to go to third order, as first order vanishes, second order is quadratic and third order gives the first odd term in $\delta a$.  The value of $\delta a$ that minimizes $(\wh{m^2}_3 -\wh{m^2}_2 -\wh{m^2}_1)$ is given by
\begin{align}
\delta a  & =  - h_{23} (2 h_{12} d_{32}-h_{23} d_{31})/d_{32}^2\, 
 \approx - s^2_{12} c^2_{12} (1-2s_{13} ) (\Delta m^2_{21})^2/ \Delta m^2_{ee} \, . 
\end{align}
This gives eq.~\ref{eq:axatm}.  The minimum of eq.~\ref{eq:321pert} reproduces the exact minimum to better than one part in $10^{8}$.

Using the above perturbation theory, one can also show that the value of $\delta a$ that minimizes $(\wh{m^2}_3 -\wh{m^2}_2)$ for NO is given by
\begin{align}
\delta a  & 
 \approx - s^2_{12} c^2_{12} (\Delta m^2_{21})^2/ \Delta m^2_{ee} \, .
\end{align}
Thus the alternative definition of the atmospheric resonance, eq.~\ref{eq:32res}, changes the corrections at ${\cal O}(s_{13} \epsilon^2)$ and similarly for IO.

\subsection{Solar Resonance}

In this Appendix we perform an perturbative expansion in the matter potential about the approximate value of the matter potential that minimizes  
$(\wh{m^2}_2 -\wh{m^2}_1)$, that is 
\begin{equation}
a= a_0 +\delta a, \quad \quad   a_0=\Delta m^2_{21} \cos 2 \theta_{12} /\cos^2 \theta_{13}
\label{eq:aRsol}
\end{equation}
Starting with the Hamiltonian from eq.~\ref{eq:ham}, a (2-3) rotation using $U_{23}(\theta_{23},\delta)$ is performed to remove the $\theta_{23}$ and $\delta$ dependence. Then a $\theta_{13}$ rotation is performed in the (1-3) sector followed by a $\pi/4$ rotation is performed in the (1-2) sector using the value of $a$ given in eq.~\ref{eq:aRsol}. The resulting Hamiltonian is given by 
\begin{align}
U^\dagger_{12}(\pi/4) U^\dagger_{13}(\theta_{13})U^\dagger _{23}(\theta_{23},\delta)~ (2E)H ~ U_{23}(\theta_{23},\delta)U_{13}(\theta_{13}) U_{12}(\pi/4)  = H_s+h
\end{align} 
where
\begin{align}
H_s =\text{\bf diag} (  &~ \Delta m^2_{21} (c^2_{12} -s_{12} c_{12}),  ~~ \Delta m^2_{21} (c^2_{12} +s_{12} c_{12}),~~\Delta m^2_{31} +a_0 ~) \, ,  \\[3mm]
& h =\frac12  \left( 
 \begin{array}{ccc}
\delta a \, c^2_{13} \quad 
 & \delta a \, c^2_{13} \quad &   \sqrt{2} \, a \, s_{13} c_{13} \\
\cdot &  \delta a \, c^2_{13} \quad  &  \sqrt{2}\, a \, s_{13} c_{13}   \\
\cdot & \cdot & 2 \delta a  \, s^2_{13}    
\end{array}  \right) \, .   \label{eq:hsol}
\end{align}
Since $h$ is symmetric, the $\cdots$'s are given by the appropriate term above the diagonal. There are only three independent elements in $h$: we will use $h_{11}, ~ h_{13}, ~h_{33}$ as the independent ones.

Now perturbation theory can be used to calculate $(\wh{m^2}_2 -\wh{m^2}_1) $, which to third order in $h$, is given by 
\begin{align}
\wh{m^2}_2 -\wh{m^2}_1
\approx  & ~d_{21} +   2h^2_{11} /d_{21}  -h^2_{13} \, d_{21}/(d_{32}d_{31})  \notag  \\
&- h^2_{13} (d_{31}+d_{32}) \,  [ \,2h_{11}/(d_{32}d_{31}d_{21}) +(h_{11}-h_{33})d_{21}/(d^2_{31}d^2_{32}) \,]
\label{eq:21pert}
\end{align} 
where  here $h_{jk}$ are the elements of matrix $h$,  eq.~\ref{eq:hsol}, and  $d_{jk}=H_s(j,j)-H_s(k,k)$, e.g. $d_{21}=\Delta m^2_{21} (2 s_{12} c_{12})$. 

Using the fact that $h_{13}$ is approximately constant,  $ h_{13} \approx a_0 \, s_{13} c_{13} / \sqrt{2} $, and  keeping  only the terms with $d_{21}$ in denominator,
then the value of $\delta a$ that minimizes $(\wh{m^2}_2 -\wh{m^2}_1)$ is approximately given by
\begin{align}
\delta a  & \approx    h^2_{13} (d_{31}+d_{32})/(d_{31} d_{32} )   
 \approx   \cos^2 2 \theta_{12} \, s^2_{13} \,  (\Delta m^2_{21})^2/ \Delta m^2_{ee} \, . 
\end{align}
This gives eq.~\ref{eq:axsol}. 
The minimum of eq.~\ref{eq:21pert} reproduces the exact minimum to one part in $10^{7}$.\\

\vspace{3cm}

\end{widetext}

\bibliography{ExFactJarlskog_v2}

\end{document}